\newcommand{\f}[1]{Fig.~\ref{#1}}
\newcommand{\eq}[1]{Eq.~(\ref{#1})}

\def\be{\begin{equation}}
\def\ee{\end{equation}}
\def\bea{\begin{eqnarray}}
\def\eea{\end{eqnarray}}
\def\l({\left(}
\def\r){\right)}

\newcommand{\y}{Y\-Ba$_{2}$\-Cu$_{3}$\-O$_{7-\delta}$}
\documentstyle[prb,aps,multicol,graphicx,psfig]{revtex}
  \renewcommand{\narrowtext}{\begin{multicols}{2}
\global\columnwidth20.5pc}
  \renewcommand{\widetext}{\end{multicols} \global\columnwidth42.5pc}
  \newcommand{\wide}{\widetext \noindent \line(200,0){245}
\line(0,1){3}\\}
  \newcommand{\narrow}{\begin{flushright}\mbox{\line(0,-1){3}$\! \!$
        \line(1,0){245}} \end{flushright} \narrowtext \noindent}

\draft{}
\begin{document}

\title 
{Superconductor strip with transport current:
Magneto-optical study of current distribution and its relaxation} 

\author{A. V. Bobyl$^{1,2}$, 
D.~V. Shantsev$^{1,2}$,
Y.~M.~Galperin$^{1,2}$, T.~H.~Johansen$^{1,}$\cite{0},
M. Baziljevich$^1$, and S. F. Karmanenko$^3$}
\address{
$^1$Department of Physics, University of Oslo, P. O. Box 1048
Blindern, 0316 Oslo, Norway\\
$^2$A. F. Ioffe Physico-Technical Institute, Polytekhnicheskaya 26,
St.Petersburg 194021, Russia\\
$^3$Electrotechnical University, Prof. Popov Str. 5, 
St. Petersburg, 197376, Russia\\
}
%\date{\today}
\maketitle

\vspace{-4.5cm}
\begin{center}
{
Submitted to Supercond. Sci. Technol. on 03.08.2001, cond-mat/0108047}
\end{center}
 
\vspace{3.5cm}

\begin{abstract} 
The dynamics of magnetic flux distributions across a
YBa$_2$Cu$_3$O$_{7-\delta}$ 
strip carrying transport current is measured using 
magneto-optical imaging at 20~K. 
The current is applied in pulses of 40-5000 ms duration and magnitude
close to the critical one, 5.5~A.
During the pulse some extra flux usually penetrates the strip,
so  the local field increases in magnitude.
When the strip is initially penetrated by flux, the local field either
increases or decreases depending both on the spatial coordinate and
the current magnitude. 
 Meanwhile, the current density always tends to redistribute
more uniformly. Despite the relaxation, all distributions
remain qualitatively similar to the Bean model predictions. 
\end{abstract}

%\pacs{PACS numbers: 74.76.Bz, 74.60.Ge, 74.60.Jg, 78.20Ls}
 
\narrowtext
\section{Introduction} 

In many technological applications like high-$T_c$ bolometers and 
electrical power cables
the superconductors are subjected to pulses of transport current.
If the current is small, the magnetic flux behavior  in
superconductors with strong bulk pinning is described by the
critical-state model (CSM).\cite{Bean} According to the model,  the
magnetic flux is frozen, and the current density $j$ is everywhere
equal to (or less than) its critical value, $j_c$. 
Flux and current distributions in superconductors carrying a transport
current were measured in several previous works. Among them
are results obtained using magneto-optical (MO)
imaging,\cite{vv1,ind1,sh1,john1,welp,pash,mocur,pss}  Hall
micro-probes,\cite{oota,herrmann,sheriff} and  detection based on THz
radiation.\cite{shikii} 

If the transport
current $I$ exceeds  its critical value $I_c$ the CSM description
becomes invalid.  In this case the flux starts to move through the
superconductor  resulting in a non-zero electric field across the
sample, i.~e., in a finite resistance.  In reality, the transition from
frozen to moving flux lattice is rather smooth, since for any value of
$I$ some regions of the  superconductor carry the current density $j
\approx j_c$.  For $I \lesssim I_c$ these regions occupy a substantial
fraction of the sample, and the magnetic flux distribution becomes
highly non-stationary.  This is known to result in a number of
dynamic effects such as voltage dependence on the current sweep rate,
and resistance relaxation.\cite{zhang,zhang99,ma,zeng}\
Direct spatially-resolved measurements of the non-stationary  flux
distribution could be very useful for analysis of these effects.
Such measurements have already proved quite efficient for the applied field case,
namely, to access the spectral distribution of the activation energies
in \y\ films,\cite{jooss} and to detect transinet vortex phases in Bi-2212
crystals.\cite{Giller}    
%However, direct spatially-resolved measurements of the non-stationary  flux
%distribution have so far been lacking.\cite{aboutGiller}  
In the present paper we use the
MO imaging technique to  study the dynamics of flux
and current distributions induced by a pulsed transport current $I$ up
to $I_c$.  The experiments are performed on a thin strip of \y\ and 
cover the time interval 40-5000~ms after
onset of the pulse.
      
It turns out that the most convenient regime to study the dynamic
effects is when the strip is subjected to both  a transport current and an
externally applied field. Then one can create states where
the CSM predicts\cite{zeld1} a region with $j=j_c$ covering more than a half 
of the strip width, and therefore the deviations from the CSM are more
easily detected. 
Such a situation has so far been studied only in a 
few works,\cite{sh1,pash,sheriff} and without
a systematic comparison to the CSM and with no attention to dynamic effects.
Therefore, we have chosen the following two magnetic states for the present study:
(i) a flux-free strip subjected to a transport current -- the reference case, and
(ii) a strip in the remanent state after a large field pulse plus a transport current. 

The paper is organized as follows. The experimental
method including the field-to-current inversion scheme is described 
in Sec.~\ref{sam}.  Section~\ref{stat} presents
comparison of the current density distributions for different $I$
to the CSM predictions. 
The relaxation of the flux and current distributions at fixed $I$
is analysed in Sec.~\ref{relax}.

\section{Experimental procedure} \label{sam}

%\subsection{Sample preparation}

Films of YBa$_2$Cu$_3$O$_{7-\delta}$ (YBCO) were grown by dc magnetron
sputtering  \cite{kar1} on LaAlO$_3$ substrate.  X-ray and Raman
spectroscopy analysis confirmed that the films were  $c$-axis oriented
and of a high structural perfection. A microbridge 
with dimensions $500\times 100 \times 0.2$ $\mu$m$^3$
was formed from the film by  a standard lithography procedure. 
Several Au wires of 50~$\mu$m diameter 
were attached to each Au/Ag contact pad using thermocompression. 

%\subsection{Magneto-optical imaging}

The imaging system is based on the Faraday rotation of
polarized light illuminating an MO-active indicator film that we
mount directly on top of the superconductor's surface. The rotated 
Faraday angle varies locally with the value of the perpendicular
magnetic 
field, and through an analyzer in an optical
microscope one can directly visualize and quantify the field
distribution across the covered sample area. As a Faraday-active
indicator we
use a  Bi-doped ferrite garnet film with in-plane
magnetization.\cite{dor1} 
The indicator film was deposited to a thickness
of 5 $\mu$m by liquid phase epitaxy on a gadolinium gallium garnet
substrate. A thin layer of aluminum is evaporated onto the
film allowing incident light to be reflected, thus providing 
double Faraday rotation of the light beam. The images were recorded
with an 8-bit Kodak DCS 420 CCD camera  and transferred to a
computer for processing. 
The MO imaging at low temperatures was performed by mounting the 
superconductor/MO-indicator on the cold finger of a continuous 
He-flow cryostat with an optical window (Microstat, Oxford).

The flux distribution in a sample biased with  large transport current
is subjected to fast relaxation.  Therefore, it was important to
minimize the camera exposure time. On the other hand, very short
exposures  lead to poor-quality MO images, and as  a compromise, 33 ms
exposure time was chosen.  The sample was biased with current pulses
of 40 ms duration, which were synchronized with the camera recording.
One image was taken during the pulse, and another image after the
current was switched off.  Experiments were also performed to study
the relaxation of flux distributions during current pulses. For this
purpose we used pulse durations up to 5000 ms. The MO images were then
recorded at different times  during the pulse.

%\subsection{Inversion}
  
As the bridge thickness
is much less than its width, the theoretical results for
the thin strip geometry\cite{zeld1,BrIn} can be used. In our
experiments, the bridge thickness is also of the order of the London
penetration depth, hence its magnetic properties are fully
characterized by the two-dimensional flux distribution at the surface.

\begin{figure}[tbp]
\centerline{\includegraphics[width=8.2cm]{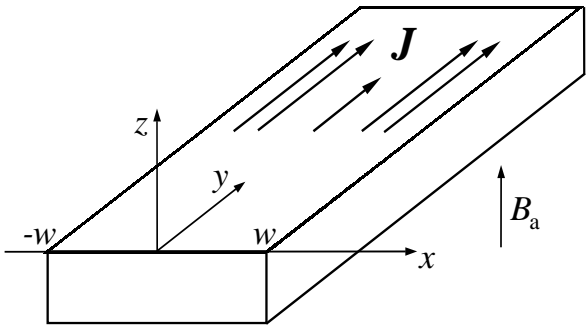}}
\caption{Superconducting strip with transport current}
\label{f_strip}
\end{figure}

The current density distribution in the strip cross section 
was calculated 
from the measured flux density distribution 
using the 1D inversion scheme proposed in Ref. \onlinecite{joh1}.
Consider a long thin superconducting strip with edges located at $x=\pm
w$,
the $y$-axis pointing along the strip, and the $z$-axis normal to the
strip
plane, see \f{f_strip}. The transport current 
flows in $y$-direction as does the screening current
due to the magnetic field,  $B_{a}$, applied along
the $z$-axis. 
The perpendicular magnetic field at the height $h$ above 
a superconductor strip, i.~e. at $-w\le x\le w$, is related to the
current
density distribution by the
Biot-Savart's law 
\begin{equation}  
B(x) = B_a + \frac{\mu_0}{2\pi} \int_{-w}^{w} dx'\,
         \frac{x'-x}{h^2+ \left( x'-x \right)^2 } \, J(x') \,.  
\label{Bh}
\end{equation} 
Here $J(x)=\int j(x,z) \, dz$ is the sheet current, $j(x,z)$ is   
the current density, and the integral is calculated over 
the strip thickness $Z$, where $Z \ll w$.
Inversion of \eq{Bh} 
yields\cite{joh1}
\begin{eqnarray} 
&&J(n)=\sum_{n'} \frac{n-n'}{\mu_0 \pi} \left( 
 \frac {1-(-1)^{n-n'}e^{\pi d}} {d^2+(n-n')^2}\right. \nonumber \\
&&\left.
 \ + 
 \frac {\left[ d^2+(n-n')^2-1 \right] \left[ 1-(-1)^{n-n'}e^{\pi
d}\right]} 
       {\left[ d^2+(n-n'+1)^2 \right] \left[ d^2+(n-n'-1)^2 \right]} 
       \right) B(n')\, . 
\label{jb} 
\end{eqnarray} 
This expression is written for discrete coordinates $x=n \Delta$ with
integer $n$, the ratio $h/\Delta$ is denoted as $d$.
The current distributions $J(x)$ calculated from this expression
are already reasonably good, see, e.~g., Refs.~\onlinecite{joh1,mocur}.
However, since the high-$k$ components are suppressed by the inversion
procedure, the accuracy of the calculated
$J(x)$ turns out to be poor near the sample edge where the current
drops abruptly to zero.  

In the present work we develop an improved inversion scheme
 using an iteration procedure.
Suppose we know $J_N(x)$ calculated on the $N$th iteration step.
We then set $J_N(x)$ equal to zero outside the superconductor,
and substitute the new current distribution into \eq{Bh} to get the
flux distribution
$B_{N}(x)$. The latter will be slightly different from
the measured distribution $B(x)$. The difference $B(x)-B_N(x)$ 
is then substituted into \eq{jb} to calculate $\delta J_N(x)$.
We then set $J_{N+1}(x) = J_N(x)+\delta J_N(x)$ and start the next
iteration.
After each iteration step $B_N(x)$ approaches closer to $B(x)$,
and the current near the edge drops more abruptly. 
It is usually sufficient to perform 5-10 iterations.
The final current distribution exhibits an abrupt jump
at the sample edges as can be seen from the figures below
and a test example discussed in Appendix A. 
In the calculations we also took into account  
effect of the component $B_x$ 
on the MO indicator sensitivity.\cite{joh1} 

\section{Pulses with different magnitude} \label{stat}

\subsection{Transport current}

The MO images for an initially flux-free 
strip carrying a transport current were taken at 20~K.
The image for the highest current, $I=4.9$~A, is shown in \f{f_image}.
The image was taken with slightly
uncrossed polarizer and analyzer.  
This provides that the flux density of different
polarity shows up as different gray level.
Consequently, the MO images can be directly converted into a 
flux density distribution using a careful calibration curve.
The flux distributions were calculated for the area marked by the black
rectangle in \f{f_image} and then averaged along the strip. 

\begin{figure} 
\centerline{\includegraphics[width=8.2cm]{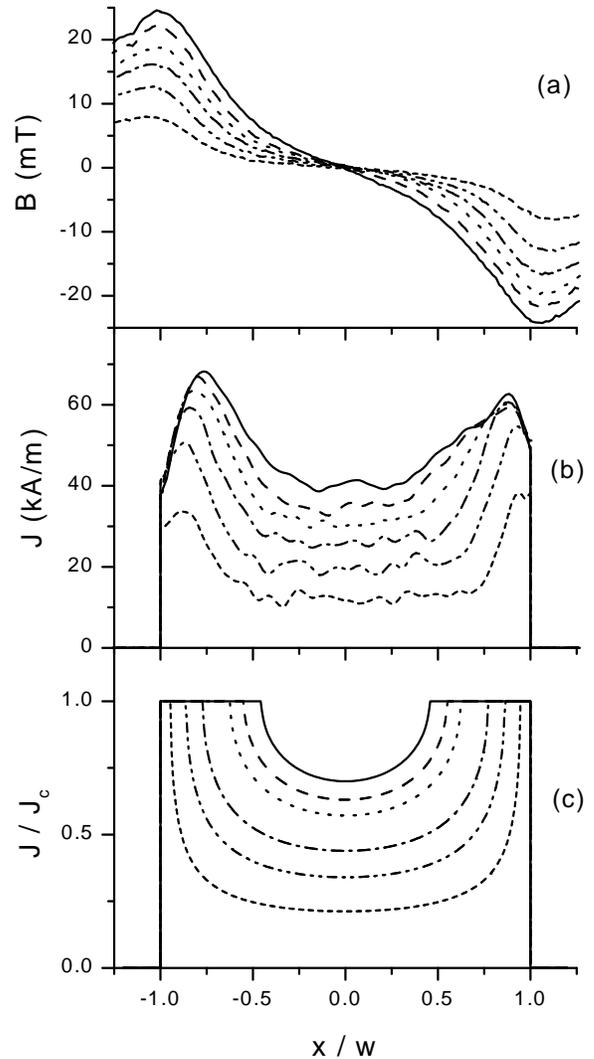}}
\caption{Initially flux-free YBCO strip carrying transport current
$I$=1.8, 2.8, 3.5, 4.3, 4.6, and 4.9 A.  
(a,b)~experimental flux and current density profiles determined from MO images
like in \f{f_image}, 
(c) the CSM current density profiles, \eq{j}, with $I_c=5.5$~A.
The flux density data were obtained from MO images of the strip,
and the plot $B(x)$ is the average profile in a 
300~$\mu$m wide band across the strip.  
\label{f_j}}
\end{figure}

Six profiles of the flux density $B(x)$ across the strip 
for different currents are shown in
Fig.~\ref{f_j}(a). The flux density has opposite signs at
the left and the right parts of the strip, and $|B|$ is maximal
near the edges. The corresponding current density profiles,
Fig.~\ref{f_j}(b),
are calculated by the inversion scheme, using $h=12 \mu$m.  
The experimental current profiles reproduce well the main features of
the Bean-model
result, Fig.~\ref{f_j}(c), where the current density is always minimal
at the center of the strip. 
On the other hand, 
the model predicts constant $J(x)=J_c$ near the edges.
Instead, experimentally we find maxima in $J(x)$ 
their magnitude and position being dependent on $I$.
This behavior is in agreement with flux creep
simulations,\cite{creepsim,mocur}
hence these deviations from the CSM can be related to 
dynamic effects.
\begin{figure} 
\centerline{\includegraphics[width=8.2cm]{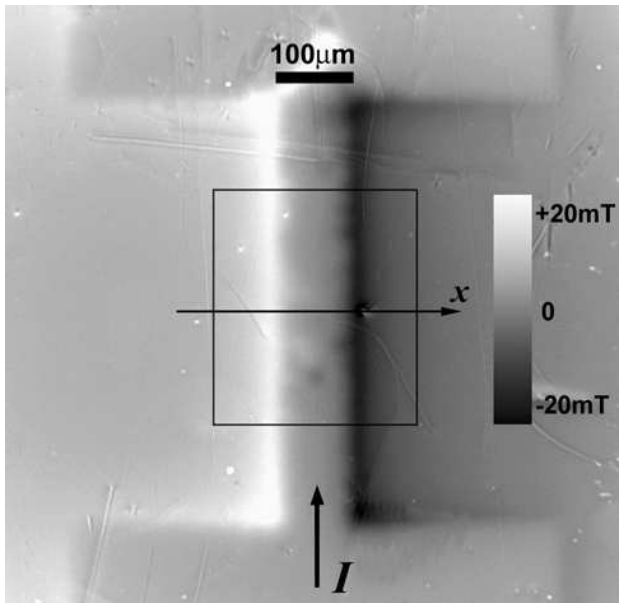}}
\caption{MO image of YBCO strip carrying 
transport current 4.9 A. The image is obtained with uncrossed 
polarizers so that the positive and negative flux density correspond to
bright and dark regions, respectively.
The black rectangle shows the image area used for calculation of
the current density distribution.  
\label{f_image}}
\end{figure}

\begin{figure} 
\centerline{\includegraphics[width=8.2cm]{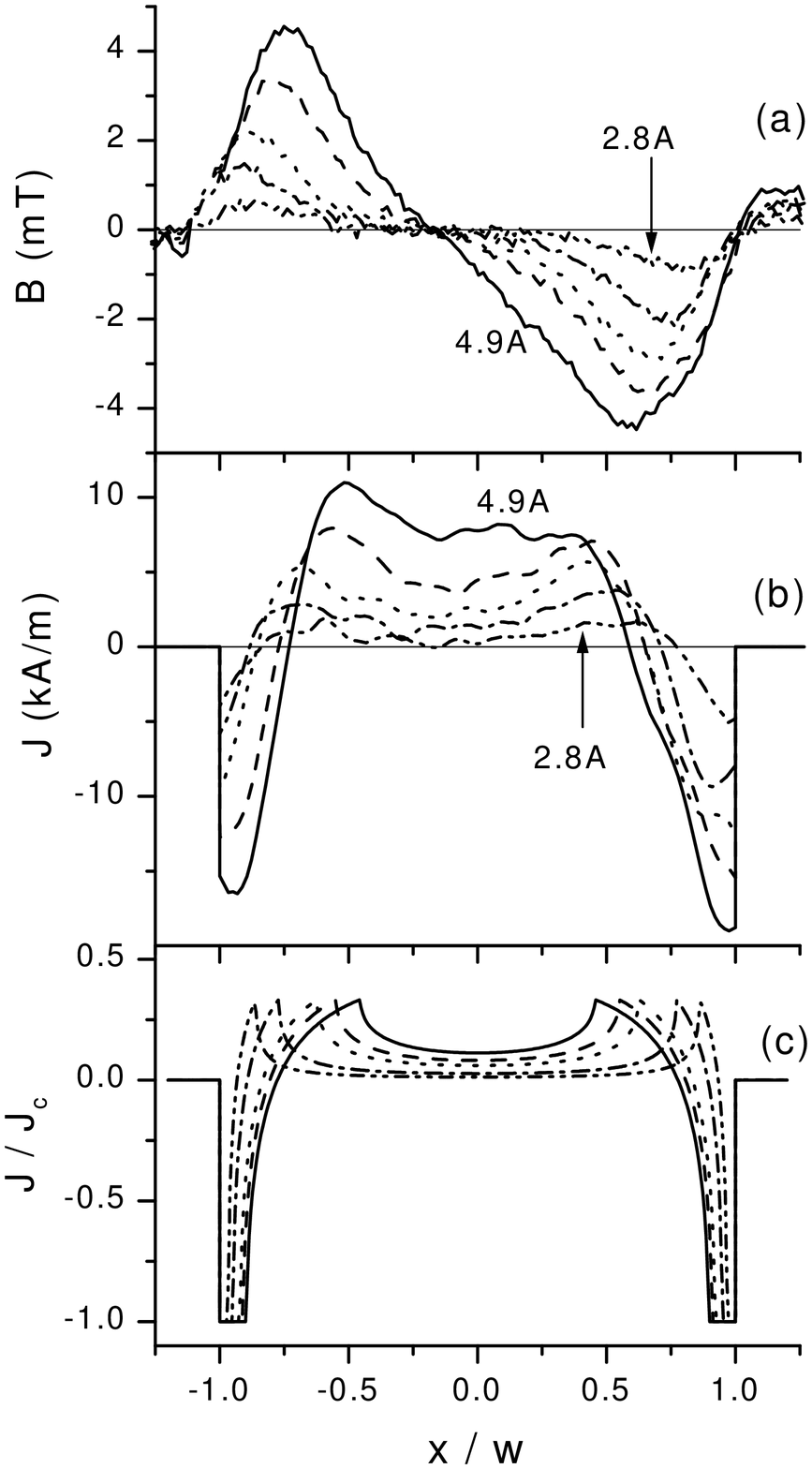}}
\caption{Remanent state of initially flux-free YBCO strip after a
  current pulse  
with $I$=2.8, 3.5, 4.3, 4.6, and 4.9 A.   
(a,b)~experimental flux and current density profiles, 
(c) the CSM current density profiles, \eq{jr} with $I_c=5.5$~A.
\label{f_jr}}
\end{figure}

When the transport current $I$ is switched off, the magnetic flux  in
the inner part of the strip remains trapped.  The return field of this
trapped flux re-magnetizes the edge regions of  the strip and 
flux of opposite sign penetrates an outer rim. 
Flux density profiles for such a remanent state 
taken across the strip are shown in
Fig.~\ref{f_jr}(a). The central part of these profiles
is similar to that of the corresponding profiles in the 
current-carrying state, Fig.~\ref{f_j}(a). However,
typical values of $|B|$ are much smaller,
in agreement with the Bean model, therefore the scale of $y$-axis is
different.    

The current density profiles  
are shown in Fig.~\ref{f_jr}(b) along with the Bean-model profiles, 
panel (c). 
As the net current must be zero in any cross section,
current lines should form closed loops. 
In the center, the current keeps flowing 
in the same direction as
the transport current, while near the edges it now flows backwards.
This result is nicely reproduced in our experiment. Moreover, 
in agreement with the Bean model we find that
(i) the current density is maximal near the edges, 
(ii) the boundary between positive and negative $J$ 
shifts inwards as $I$ increases.
Nevertheless, like in the current-carrying state, we don't find
an agreement with the theory near the edges where $J(x)$
should be always equal to $J_c$.

The flux dynamics is expected to proceed much faster in the regions
where the current density is close to the critical one. It is exactly
those edgy regions where the deviations from the Bean behavior are
found.
However, in the above experiment, the size of these regions
was relatively small, $0.6-0.9w<|x|<w$. 
Thus, the results could possibly be affected by inaccuracy of the
inversion
method or small defects near the strip edge.
For this reason we performed another experiment where the region
of $j \approx j_c$ is much more extended, and unambiguous manifestations
of the dynamic effects are found.

\begin{figure} 
\centerline{\includegraphics[width=8.2cm]{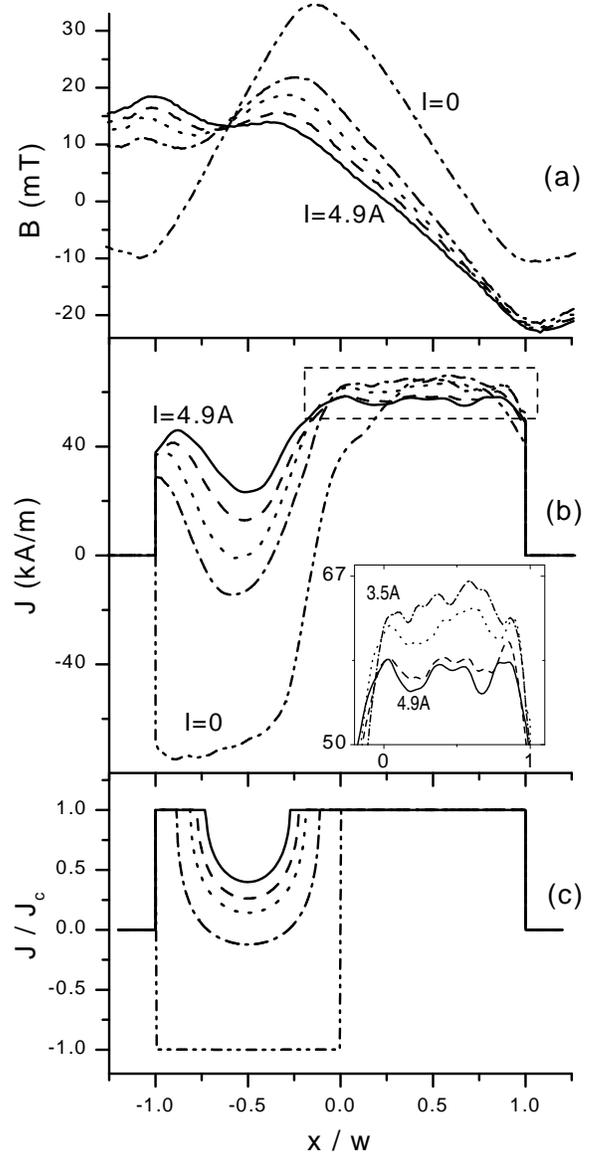}}
\caption{YBCO strip, initially in fully-penetrated state, carrying
  transport current 
$I$=0, 3.5, 4.3, 4.6, and 4.9 A.  
(a,b)~experimental flux and current density profiles, (c) the CSM
current density profiles, \eq{rb}, with $I_c=5.5$~A.
Inset in (b) is a blow-up of the dashed rectangle.
\label{f_rb}}
\end{figure}

\subsection{Remanent Field + Transport current}

In the second experiment, a large magnetic field was applied to the
strip
and subsequently removed.
As a result, some flux remain trapped in the strip, and the flux profile
has a conventional triangular shape with maximum
at the center, see the ``$I=0$'' curve in \f{f_rb}(a).
Small regions near the edges, $|x|>w/\sqrt{2}$ are filled with return
field
of the opposite polarity. 
The strip was then biased with transport current $I$, and flux density
distributions were measured in the current-carrying state.
$B(x)$ profiles across the strip for different $I$ are shown in
\f{f_rb}(a).
As the current increases, the central peak  diminishes, 
and almost disappears at the maximal current (solid line).
Meanwhile, another peak at the left edge shows up and grows with $I$.
At intermediate currents one can clearly see two peaks in $B(x)$
profile.
This feature was predicted long ago within the Bean model
analysis,\cite{zeld1}
but has not so far been observed in experiment.

\begin{figure} 
\centerline{\includegraphics[width=8.2cm]{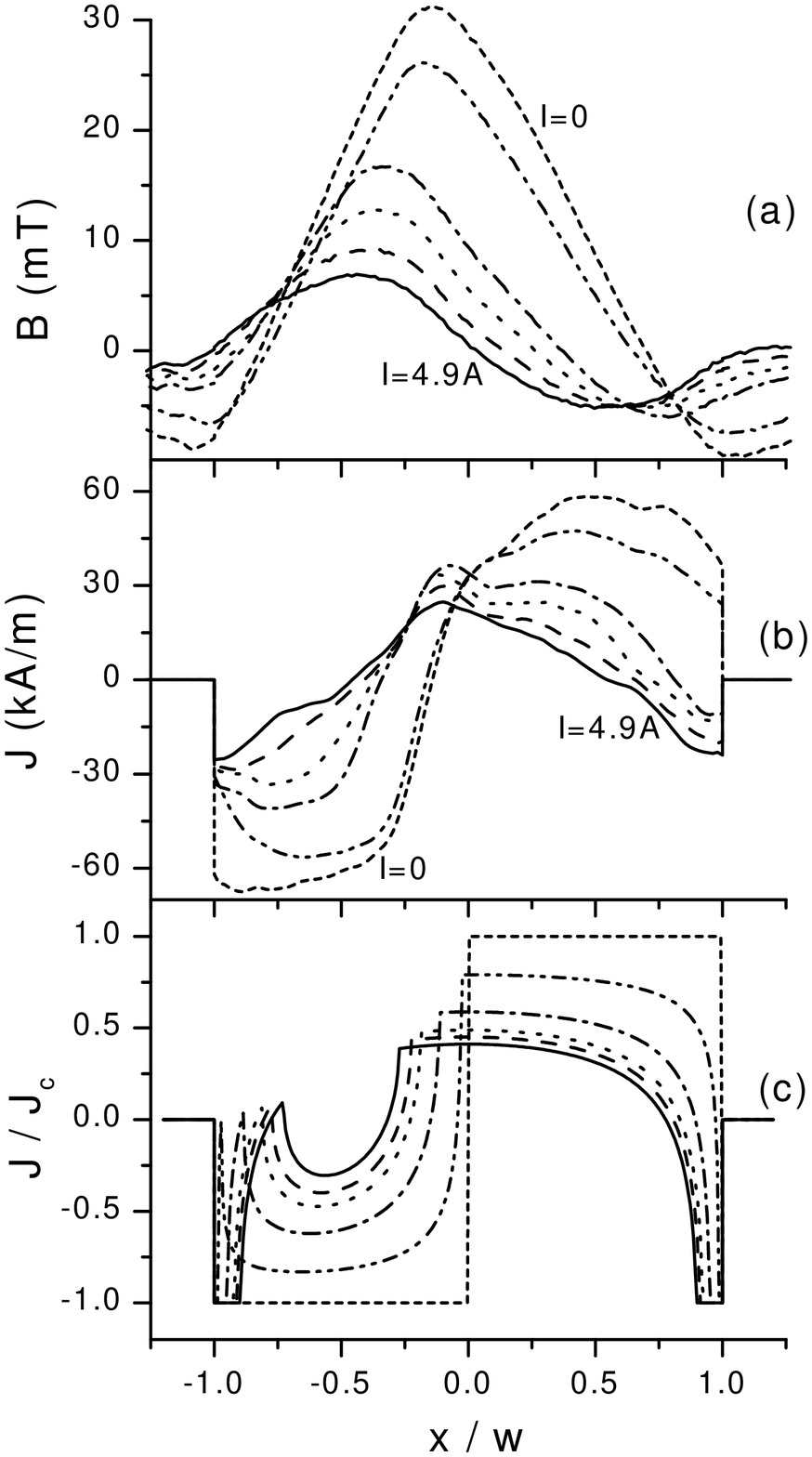}}
\caption{Remanent state of initially fully-penetrated YBCO strip after a
current pulse
$I$=0, 1.8, 3.5, 4.3, 4.6, and 4.9 A.  
(a,b)~experimental flux and current density profiles, (c) the CSM
current density profiles, \eq{rbr}, with $I_c=5.5$~A.
\label{f_rbr}}
\end{figure}

The current density profiles
are presented in \f{f_rb}(b) along with the Bean model profiles, panel
(c). 
According to the model, the current distribution in the left half is 
actually a copy of the distribution for the ``pure'' transport
current state, \f{f_j}(c), though squeezed in $x$-direction. 
Thus, $J(x)$ profile has a deep
minimum at the middle of the left half, which is 
well reproduced in the experimental profiles. 
However, we find a clear disagreement with the theory 
in the right half of the strip.
There the Bean model predicts a flat $J(x)$ 
unaffected by the transport current. In experiment, $J(x)$,
though being quite flat appears dependent on $I$.
This is emphasized in the inset,  where one can see a monotonous
{\em decrease} of $J(x)$ level as
 the current {\em increases}.
Such a behavior contradicts not only the Bean model but also
common intuition. 
This interesting observation is
discussed below in the framework of relaxation effects.

The remanent flux and current distributions after the current pulse 
are shown in \f{f_rbr}.
One can clearly see a current-induced suppression and shift
of the peak in $B(x)$, as predicted by the Bean model,\cite{zeld1}
and observed earlier.\cite{sheriff} 
However, the current density profile is too complicated to make any 
solid conclusions. We mention only two main features
of the Bean profiles found also experimentally: 
(i) a peak in $J(x)$ 
at small negative $x$, (ii) a concave shape of the profile in the left,
and convex in right half of the strip.

\section{Relaxation during the pulse}\label{relax}

In order to understand the reason for the observed deviations 
of the measured $B$- and $J$-profiles from the CSM predictions
we analyze the effect of pulse duration.
It was found that the flux distribution does not remain fixed
during the current pulse, even though the current magnitude is kept
constant. The flux is subjected to relaxation which is especially fast
for larger currents.

The relaxation effects are known to be noticeable in high-temperature
superconductors even  
in the absence of transport current due to large flux creep
rates.\cite{yeshurun}
The direction in which the relaxation proceeds can 
be always well understood. In a conventional experiment, the flux
trapped
in a superconductor kept at 
zero applied field monotonously decreases with time. 
This means that the flux density decreases in amplitude
throughout the sample. 
If a sample is kept in constant non-zero applied field $H_a$,
then the flux distribution relaxes towards a uniform $H(\vec{r})=H_a$.
Remarkably, for a thin sample in a perpendicular field there
coexist regions with $H>H_a$, where $H$ decreases 
due to creep, and regions with $H<H_a$, where $H$
increases.\cite{neutral,neutral-ex}
They are separated by a so-called neutral line, where the flux density
remains constant.

\begin{figure} 
\centerline{\includegraphics[width=8.2cm]{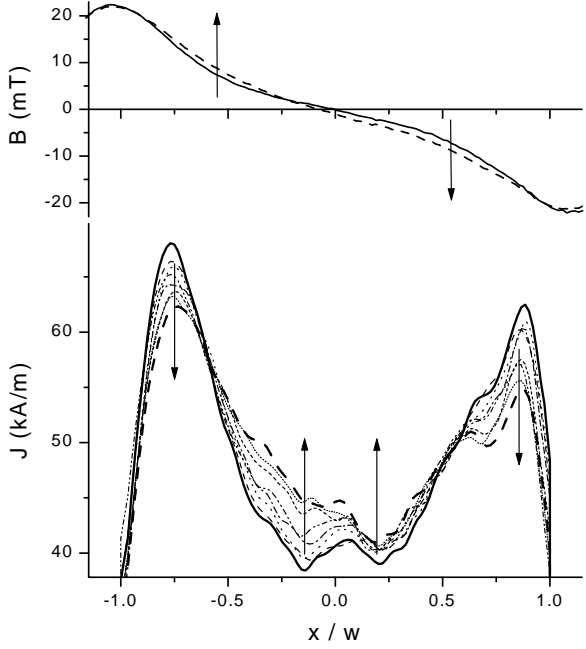}}
\caption{Relaxation of flux and current density distribution
in an initially flux-free YBCO strip 
during the current pulse, $I=4.9$~A. The profiles correspond to
different times:
40, 100, 200, 300, 450, 800, 1200, and 1600~ms; arrows show the time 
direction. The absolute flux density increases throughout the sample.  
The current tends to redistribute more uniformly: some extra 
current moves from the edges to the central region.  
\label{f_relax}}
\end{figure}

\begin{figure} 
\centerline{\includegraphics[width=8.2cm]{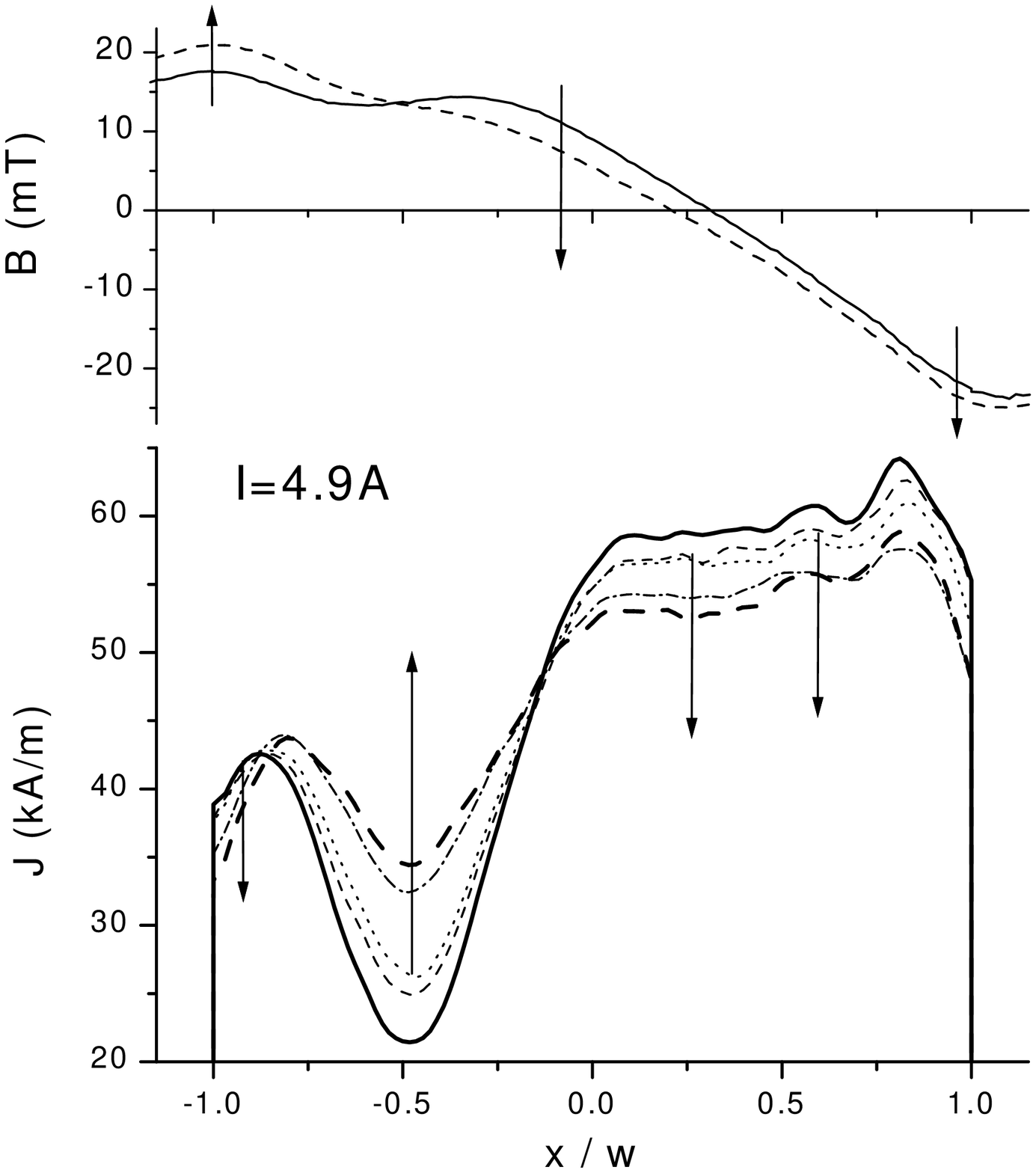}}
\centerline{\includegraphics[width=8.2cm]{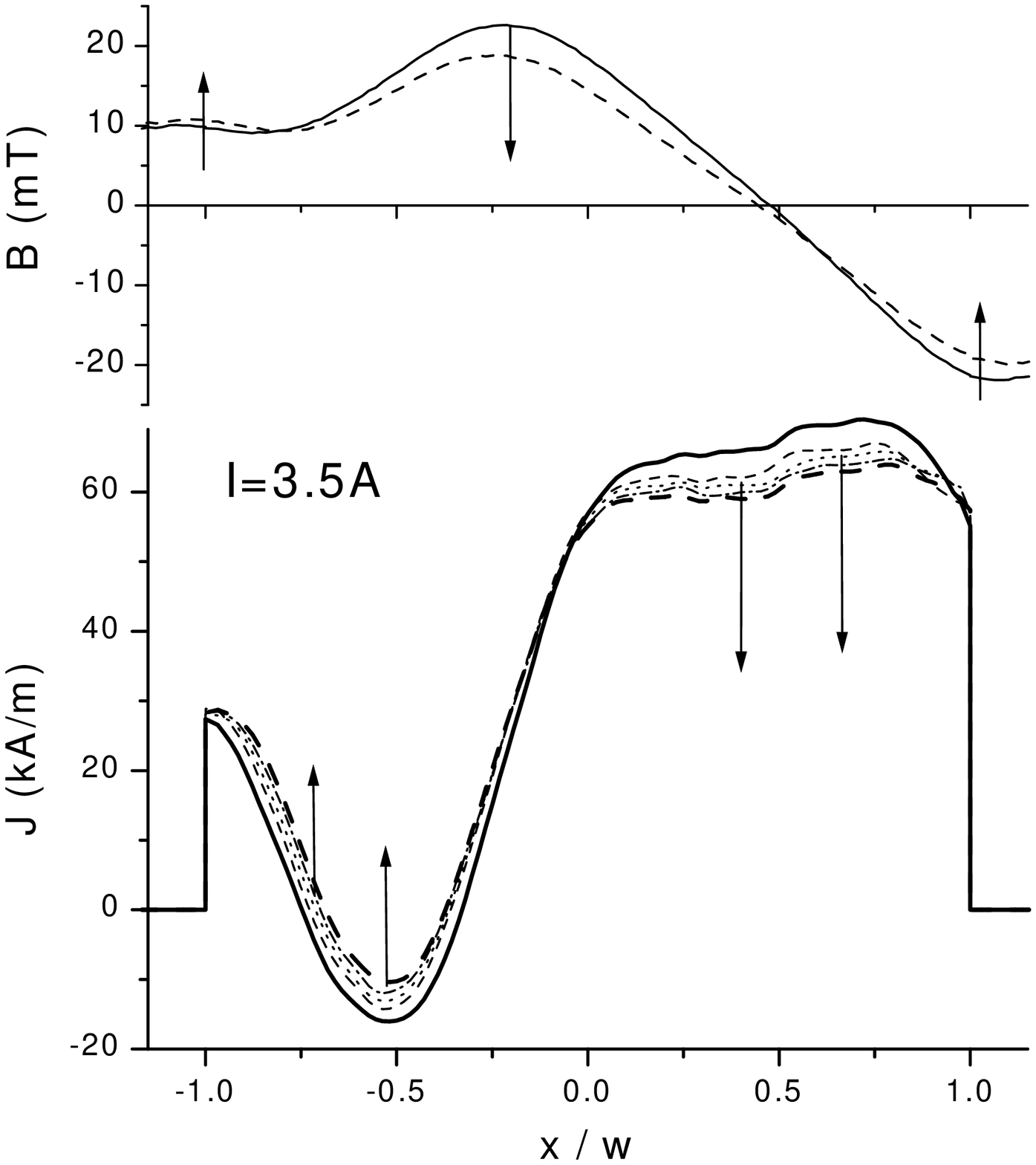}}
\caption{Relaxation of flux and current density distribution during the
current pulse
for the ``remanent field + current'' state.
Top: $I=4.9$~A, times: 40, 100, 200, 800 and 1600~ms. 
Bottom: $I=3.5$~A, times: 40, 400, 800, 1600, and~5000 ms. 
Arrows show the time 
direction. The current tends to redistribute more uniformly: some extra 
current moves from the high-current right half to the low-current
left half.  
\label{f_relax2}}
\end{figure}

In the presence of transport current the relaxation
of flux and current density profiles is even more  
complicated as illustrated by Figs.~\ref{f_relax},
and~\ref{f_relax2}.
Flux distributions were measured at different times during
the pulse starting from $t=40$~ms up to 5000~ms
($t=0$ corresponds to switching current on).
\f{f_relax} shows the results for
the case of initially flux-free strip with $I=4.9$~A, like in \f{f_j}.
Only the first ($t=40$~ms) and the last ($t=1600$~ms) $B(x)$ profiles
are
presented. The two corresponding $J(x)$ profiles are
indicated by thick solid and thick dashed line respectively
at the bottom panel.

We note that, in contrast to
the applied field case, the relaxation leads to {\em increasing}
the field amplitude throughout the superconductor.
This means that more and more flux enters the sample as
the time passes.
In terms of $J$-profile it means that
some extra current redistributes
from the edges to the center. 
It can be also interpreted as a tendency for $J$-distribution
to become more uniform.

Shown in \f{f_relax2} are results 
for the combined ``remanent field + transport current" state, with
$I=3.5$ and~$4.9$~A.
Here the relaxation of $B(x)$ does not follow a simple rule.
The flux density may increase or decrease depending on the position in
the sample,
and on the current amplitude. Moreover, a closer look shows that 
the direction of relaxation may change as the relaxation proceeds, like
in \f{f_relax2}(a) at $x\approx -0.8w$.
However, the changes in $J$-profile have the same tendency
as for the flux-free strip subjected to transport current: 
the current tends to distribute more uniformly.

In Refs.~\onlinecite{mcelf,darwin},
the relaxation of the local flux density measured by
a Hall sensor was described 
using an ``effective'' CSM with a \emph{time-dependent} critical current density.
This description is basically in agreement with our results.
Indeed, when $J_c$ decays with time, the CSM predicts more and more
uniform $J(x)$-distribution until at some point one comes to 
a completely uniform $J(x)=J_c(t)$.
The rate of the $J_c$ decay can be estimated from the plateau  
in $J(x)$-profiles at $x>0$ in \f{f_relax2}.
We find that from $t=40$ to $5000$~ms the $J_c$ reduction was $\approx
10$\%.

We have also performed computer simulations of
flux creep based on the Maxwell equation,
and Bio-Savart's law for a thin strip, \eq{Bh}, and logarithmic current
dependence of the pinning energy.\cite{mocur,creepsim} The direction in 
which  the relaxation of flux and current proceeds
was always in agreement with our experimental results.
The same tendency has been also found in simulations reported in
Ref.~\onlinecite{zeng}.

However,  the creep simulations do not explain the decrease of $J$
with transport current  
increase illustrated in the inset of Fig.~\ref{f_rb}.
One can see from comparison of the top and bottom panels of 
\f{f_relax2} that this apparent decrease
is related to a faster relaxation of $J$-distribution for larger $I$.
Consequently, the relaxation rate of $J_c$ turns out to be
$I$-dependent.
The explanation of this effect remains an open question.
One can expect that at larger currents 
a slow permanent flow of vortices across the strip sets in, probably
along a few channels. This permanent flow does not change 
the global flux distribution and cannot be resolved magneto-optically.
However, it keeps vortices in motion which should 
reduce $J_c$ and assist relaxation processes.

\section{Conclusion} 
 
We have studied the dynamics of flux and current density distributions 
in a YBCO strip subjected to a pulse of transport current at 20 K.
Relaxation of the distributions during the pulse is much different
from the conventional relaxation of trapped flux in the absence of
current. The local flux density may decrease or increase depending
on the position in the sample, current amplitude, and magnetic history.
However, the relaxation of $J$-distribution always has the same
tendency: the current tends to redistribute more uniformly.
Despite the relaxation, most flux and current distributions remain
in a qualitative agreement with the Bean model predictions and can
be described by a Bean model with a time-dependent $J_c$.
However the behavior for the current-carrying strip in initially
fully-penetrated state is not fully understood. 
This behavior is suggested to be associated with
a flux-flow induced reduction of $J_c$.    

\acknowledgements

The financial support from the Research Council of Norway, from the
RFBR ``Scientific Schools'' project 00-15-96812, and 
Russian Program for Superconductivity, project No 98031, 
is gratefully acknowledged. We are grateful to Bj{\o}rn Berling for a 
many-sided help.

\appendix
\section{ The inversion method}

We demonstrate here advantages of the proposed iteration inversion
method
for calculation of the current density profile $J(x)$
across the strip from the measured field profile.
For this purpose we use a model example of a strip
with Bean-model $J(x)$ given by \eq{j}, with $a=0.5w$.
The corresponding flux density profile $B(x)$
at the height $h$ above the strip is given by \eq{Bh}.
To model our experimental situation we assume that
$B(x)$ is measured 
in the range $-4w<x<4w$ at fixed $x$ separated by $\Delta=0.02w$, and
that $h=0.2w$.

First, we restore the current density profile using
the standard procedure based on \eq{jb}.
Straightforward calculations where $d=h/\Delta=10$ produce
highly oscillating $J(x)$ due to
contribution of higher harmonics, see \onlinecite{joh1}.
The best results are obtained
by the choice $d=2$, which means that only  
$B$-values at $x$ separated by $h/2$
are used. This $J(x)$ profile shown by
the dashed line in  \f{f_test} reproduces qualitatively
the initial $J$-profile (dotted line). 
There are, however, large distortions near the edges,
and an error in the overall current magnitude.

The solid line shows the $J$-profile obtained by the 
iteration procedure described in Sec. II.C, for 5 iterations.
It is clear from the figure that the iteration method
is more successful in restoring the correct $J$-profile.
It should be noted that the exact form of \eq{jb} is not essential
for the iteration procedure. It can be modified if it
will help improve convergence of $B_N(x)$ towards $B(x)$.
The best convergence is found when parameter $d$ is equal to unity or
less.

\begin{figure} 
\centerline{\includegraphics[width=8.2cm]{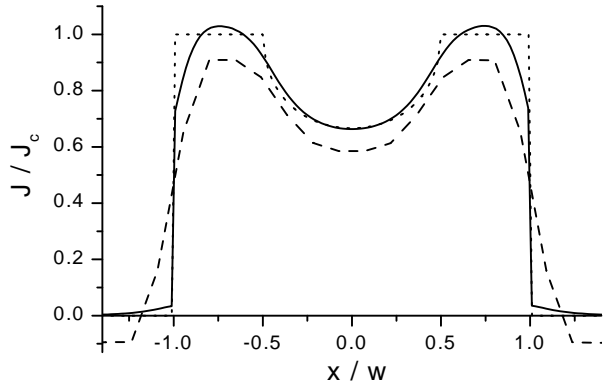}}
\caption{Current density distributions 
obtained by iterative inversion procedure (solid line),
and standard procedure (dashed line) for a Bean-model field 
distribution. The dotted line is Bean-model current distribution.  
\label{f_test}}
\end{figure}

The advantage of the iteration procedure is explained by the fact 
that the inversion formula, \eq{jb},
is used to calculate only $\delta J_N(x)$ instead of the whole $J(x)$.
Normally, $\delta J_N(x)$ is much smaller and smoother than $J_N(x)$.
Thus, an inevitable error intrinsic to any inversion procedure is
minimized. 
The only assumption used by this
procedure is that current cannot flow outside the superconductor sample,
which is physically well-grounded. The location of
the sample edges can be determined with high accuracy from
a simple optical image.

\section{Bean-model results}

Here we list some Bean-model expressions
for current density distributions in  
a thin superconducting strip obtained in Refs.~\onlinecite{zeld1}
and~\onlinecite{BrIn}. 
For an initially flux-free strip $-w \le x \le w$, Fig.~\ref{f_strip}, 
carrying transport current $I$
\wide
\begin{equation}
\frac{J(x)}{J_c} = \left\{ \begin{array}{lr}\frac{2}{\pi} \arctan
\left(\sqrt{\frac{w^2-a^2}{a^2-x^2}}\right), & |x|<a \\     
 1, & a<|x|<w 
              \end{array} 
       \right. , 
\label{j} 
\end{equation}
where $a=w\sqrt{1-(I/I_c)^2}$, and $I_c=2w\, J_c$ is the critical
current. This distribution is shown in \f{f_j}(c).
In the remanent state, after the transport current is switched off, 
\begin{equation}
\frac{J(x)}{J_c} = \left\{ \begin{array}{lc}
  \frac{2}{\pi}\left[ 
              \arctan \left(\sqrt{\frac{w^2-a^2}{a^2-x^2}}\right) - 
              2 \arctan 
              \left(\sqrt{\frac{w^2-b^2}{b^2-x^2}}\right) 
              \right] , & -a<x<a, \\
 1- \frac{4}{\pi}\arctan 
              \left(\sqrt{\frac{w^2-b^2}{b^2-x^2}}\right) , & a<|x|<b 
              \end{array} 
       \right.\, , 
\label{jr} 
\end{equation}
where $a=w\sqrt{1-(I/I_c)^2}, \  b=w\sqrt{1-(I/2I_c)^2}$.
At $b < |x| <w $ the   current density is equal to $-J_c$, see
\f{f_jr}(c). 

When a large magnetic field was applied to the strip
and subsequently removed, the current density $J(x)=-J_c$
and $J(x)=J_c$ in the left and the right half of the strip,
respectively.
If the strip is subsequently biased with transport current $I$, then  
\begin{equation}
  \frac{J(x)}{J_c} = \left\{ \begin{array}{lc} 
      \frac{4}{\pi}\arctan 
              \left(\sqrt{\frac{(w/2)^2-a^2}{a^2-(x+w/2)^2}}\right) - 
              1  , & -w/2-a<x<-w/2+a, \\ 
               1 , & -w<x\le-w/2-a {\rm ~~or~} -w/2+a\le x <w 
              \end{array} 
       \right.\, , 
\label{rb} 
\end{equation}
see \f{f_rb}(c).
In the remanent state after the current pulse, see \f{f_rbr}(c),
\begin{equation}
 \frac{J(x)}{J_c}  = \left\{ \begin{array}{lc} 
 -1-\frac{4}{\pi}\arctan 
              \left(\sqrt{\frac{w^2-b^2}{b^2-x^2}}
              \right) + \frac{4}{\pi}
              \arctan\left(\sqrt{\frac{(w/2)^2-a^2}{a^2-(x+w/2)^2}} 
\right) , & -\frac{w}{2}-a {\rm ~~or~} -\frac{w}{2}+a\le x <b, \\ 
                                                                  -1 ,
    & b \le |x| <w  
              \end{array} 
       \right.\,  
\label{rbr} 
\end{equation}
\narrow

\widetext 
\end{document}